\documentstyle[aps,manuscript]{revtex}

\makeatletter

\newcommand{\LyX}{L\kern-.1667em\lower.25em\hbox{Y}\kern-.125emX\@}

\makeatother

\begin{document}

\title{Scattering map for two black holes}

\author{Alessandro P. S. de Moura\thanks{
email: sandro@ifi.unicamp.br
}}

\address{Instituto de F\'{\i}sica Gleb Wataghin, UNICAMP, 13083-970 Campinas SP, Brazil}

\author{Patricio S. Letelier\thanks{
email: letelier@ime.unicamp.br
}}

\address{Instituto de Matem\'{a}tica, Estat\'{\i}stica e Ci\^{e}ncia da Computa\c{c}\~{a}o,
Departamento de Matem\'{a}tica Aplicada, UNICAMP, 13083-9790 Campinas SP, Brazil}

\maketitle
\begin{abstract}
We study the motion of light in the gravitational field of two Schwarzschild
black holes, making the approximation that they are far apart, so that the motion
of light rays in the neighborhood of one black hole can be considered to be
the result of the action of each black hole separately. Using this approximation,
the dynamics is reduced to a 2-dimensional map, which we study both numerically
and analytically. The map is found to be chaotic, with a fractal basin boundary
separating the possible outcomes of the orbits (escape or falling into one of
the black holes). In the limit of large separation distances, the basin boundary
becomes a self-similar Cantor set, and we find that the box-counting dimension
decays slowly with the separation distance, following a logarithmic decay law. 
\end{abstract}

\section{Introduction}

In this article, we study the motion of light (null geodesics) in the gravitational
field of two non-rotating Schwarzschild black holes. In general relativity,
solutions of the field equations describing more than one purely gravitational
sources are necessarily non-stationary, because gravity is always attractive
(we are not considering exotic matter); there is no possibility of arbitrarily
``pinning'' sources as is done in Newtonian gravitation, because of the automatic
self-consistency of the nonlinear Einstein's equations. If we demand that the
two black holes be fixed in space, then the solution includes a conical singularity
(a ``strut'') lying on the axis on which the two masses are located\cite{kramer}.
This singularity appears as a natural consequence of the field equations, and
it is necessary to keep the two masses from falling towards each other. However,
this singularity would have to be made of very exotic matter, and this solution
does not describe any realistic system in astrophysics. The real solution to
the relativistic two-body problem can have no conical singularities, and it
is necessarily non-stationary. The two black holes will spiral around each other,
emitting gravitational waves, which makes this problem even more difficult.
There is no exact solution for the relativistic two-body problem, and even a
numerical solution has eluded the most powerful computers.

In order to cope with this problem, Contopoulos and others \cite{contopoulos,cornish}
have used the Majumdar-Papetrou solution \cite{kramer} to study the dynamics
of test particles in a space-time with two black holes. The Majumdar-Papetrou
metric used by Contopoulos describes two non-rotating black holes with extreme
electric charge (\( Q=M \) in relativistic units), whose gravitational pull
is exactly matched by their electrostatic repulsion, thereby allowing a static
mass configuration. They have found that in this metric the motion of both light
and massive particles is chaotic, with a fractal invariant set and a fractal
basin boundary. However, it is very unlikely that the Majumdar-Papetrou metric
describes realistic astronomical objects, since there is no known realistic
astrophysical process by which a black hole with extreme charge could be formed.
Even though the two black holes with extreme charge have proven useful as toy
models, it is important to address the more realistic problem of two uncharged
black holes. This is what we do in this article, for the motion of light and
other massless particles.

In order to overcome the fact that there is no static solution for the two-black-hole
system, we consider the case when the two black holes are far apart, with a
distance much larger than their Schwarzschild radii. In this case, nonlinear
effects in the field equations are expected to be small, and we can approximate
the motion of test particles in the neighborhood of one of the masses as being
the result of the field of that mass alone, and disregard the effect of the
other black hole as being negligible. Using this approximation, the motion of
test particles in the two-black-hole system is treated as a combination of motions
caused by isolated Schwarzschild black holes. Since the equations of motion
for the Schwarzschild geometry can be analytically integrated, our dynamical
system is reduced to a map, which is much easier to study than a system of ordinary
differential equations. This scattering map is built in Section 2, for the simple
case of two black holes with equal masses. In Section 3, we show that this map
has a fractal basin boundary separating the possible outcomes of a light ray
in the two-black-hole field, namely, falling into either of the black holes
or escaping towards the asymptotically plane infinity. The fractal (box-counting)
dimension of this basin boundary is numerically calculated, and the sensitivity
to initial conditions implied by the fractal nature of the boundary is thereby
quantified. In Section 4 we use explicitly the condition of large separation
between the black holes. In this limit, the basin boundary becomes a self-similar
Cantor set, which allows us to obtain some analytical results. One of our main
results is that the fractal dimension decays very slowly (logarithmically) with
distance. The slow decay of the fractal dimension makes it more likely that
the fractal nature of the basin boundary has some importance for astrophysics.
In section 5, we consider the case of two black holes with unequal masses, in
the limit of a large separation; we find that the logarithmic decay law of the
fractal dimension for large distances is also valid in this case. In section
6, we summarize our results and draw some conclusions.

\section{Scattering map for two black holes with equal masses}

We begin by reviewing some basic results concerning the motion of test particles
in the field of an isolated Schwarzschild black hole\cite{chandra,frolov}.
We consider specifically the case of null geodesics, which concerns us most,
but many features of the dynamics also apply to massive particles.

The Schwarzschild metric is written in spherical coordinates as:
\[
ds^{2}=\left( 1-\frac{2M}{r}\right) dt^{2}-\frac{dr^{2}}{1-\frac{2M}{r}}-r^{2}d\Omega ^{2},\]
with \( d\Omega ^{2}=d\theta ^{2}+\sin ^{2}\theta d\phi ^{2} \) being the element
of unit area, and \( t \) is the time measured from a distant observer. \( M \)
is the black hole's mass in geometrized units. We are interested only in the
region of space-time outside the event horizon, \( r>2M \). Due to the conservation
of angular momentum, test particles move on a plane, which can be chosen as
\( \theta =\pi /2 \). The plane whereon the motion occurs is then described
by the coordinates \( r \) and \( \phi  \). The geodesic equations which describe
trajectories of test particles on this plane can be analytically integrated
by means of elliptical functions\cite{chandra}. Here we are interested in the
scattering of null geodesics by the black hole. A light ray coming from infinity
towards the black hole is characterized by the impact parameter \( b \) defined
by the ratio \( b=L/E \), where the angular momentum \( L \) and the energy
\( E \) are constants of motion given by:
\[
E=\left( 1-\frac{2M}{r}\right) \frac{dt}{d\lambda };\]
\[
L=r^{2}\frac{d\phi }{d\lambda };\]
\( \lambda  \) is the geodesic's affine parameter. For null geodesics only
the ratio of \( L \) and \( E \) is of importance to the dynamics. In the
asymptotically plane region \( r\rightarrow \infty  \), \( b \) corresponds
to the usual impact parameter of classical scattering problems.

If the impact parameter is below the critical value \( b_{c}=3\sqrt{3}M \),
the trajectory of the light ray spirals down the event horizon and plunges into
the black hole. If \( b>b_{c} \), the trajectory circles the black hole and
escapes again towards infinity, being deflected by an angle \( \Delta  \).
The lowest value \( P \) of the radial coordinate \( r \) along the trajectory
(the ``perihelium'') is given by:
\begin{equation}
\label{eqP}
b^{2}=\frac{P^{3}}{P-2M}.
\end{equation}
Following Chandrasekhar\cite{chandra}, we define the quantities \( Q \), \( k \)
and \( \chi  \) by:
\begin{equation}
\label{eqQ}
Q^{2}=(P-2M)(P+6M);
\end{equation}
\begin{equation}
\label{eqk}
k^{2}=\frac{Q-P+6M}{2Q};
\end{equation}
\begin{equation}
\label{eqchi}
\sin ^{2}\left( \chi /2\right) =\frac{Q-P+2M}{Q-P+6M}.
\end{equation}
The scattering angle \( \Delta  \) is then given by
\begin{equation}
\label{eqDelta}
\Delta =\pi -2f(b),
\end{equation}
and the function \( f(b) \) is:
\begin{equation}
\label{eqf}
f(b)=2\sqrt{\frac{P}{Q}}\left[ K(k)-F(\chi /2,k)\right] .
\end{equation}
Here \( F \) is the Jacobian elliptic integral and \( K \) is the complete
elliptic integral. In Fig. 1 we show a plot of \( \Delta (b) \). As \( b \)
approaches the critical value \( b_{c} \) from above, \( \Delta  \) goes to
infinity; trajectories with \( b \) sufficiently near \( b_{c} \) can circle
the black hole an arbitrary number of times before escaping, and for \( b=b_{c} \),
the light ray makes an infinite number of rotations, and never escapes. This
is a consequence of the existence of an unstable periodic orbit at \( r=3M \),
which appears as a maximum in the effective potential. The orbits with \( b=b_{c} \)
spiral towards the \( r=3M \) orbit, and in the language of dynamical systems
they make up the stable manifold associated with this periodic orbit. 

The fact that \( \Delta  \) assumes values above \( \pi  \) for a non-zero
range of \( b \) implies the existence of a rainbow singularity in the scattering
cross section; this is to be contrasted with the Newtonian Rutherford scattering,
which shows no such singularities. In fact, \( \Delta  \) assumes arbitrarily
large values, and the differential cross section at any given angle \( \theta  \)
is made up by an infinite number of contributions arising from trajectories
with \( \Delta =\theta  \), \( \Delta =\theta +2\pi  \), in general \( \Delta =\theta +2n\pi  \),
corresponding to trajectories that circle the black hole \( n \) times before
being scattered towards \( \theta  \). However, large values of \( n \) correspond
to very low ranges of \( b \): the set of trajectories that scatters by \( \theta +2n\pi  \)
has a measure that decreases very rapidly with \( n \). Chandrasekhar \cite{chandra}
shows that the impact parameter \( b_{n} \) corresponding to a scattering by
\( \theta +2n\pi  \) for large values of \( n \) is given approximately by:
\begin{equation}
\label{expon}
b_{n}=b_{c}+3.48Me^{-(\theta +2n\pi )}.
\end{equation}
This expression shows that the measure of the set of trajectories scattered
by \( \theta +2n\pi  \) decays exponentially with \( n \), and the contribution
of orbits with large \( n \) to the cross section is small. In fact, we shall
see later that in many cases it is a good approximation to consider only orbits
with \( n=0 \).

After reviewing some properties of an isolated black hole, we now consider the
case of two black holes with equal mass \( M \) (we consider the case of different
masses in Section 5). As we mentioned in the introduction, there is no exact
solution of Einstein's field equations that describes this system. Because of
this, we assume that the two black holes are separated by a distance \( D \)
much larger than their Schwarzschild radius \( 2M \); in this limit the nonlinear
interaction between the two gravitational fields can be ignored. In a real system,
the two black holes will be rotating around their center of mass; however, their
rotation speed is much smaller than the velocity of light. We can thus consider
the two black holes to be fixed in space, without incurring in too much error.
Notice that this approximation might not be valid for massive test particles.

We are interested in the orbits that never escape to infinity nor fall into
one of the event horizons; these orbits make up the basin boundary of the system,
which will be discussed later in more detail. For the orbits not to escape,
they need to have impact parameters such that they are scattered by at least
\( \pi  \) by one of the black holes. In the case of an isolated black hole,
this corresponds to an impact parameter lower than \( b=b_{esc}\approx 5.35696M \),
which is less that three times the Schwarzschild radius. Since in our approximation
\( D\gg 2M \), for the purpose of finding the basin boundary we can consider
that the light rays are scattered by each black hole separately, the other black
hole being too far away to make a significant difference in the scattering.
After suffering a scattering by one of the black holes, the ray may reach the
other black hole, depending on its emerging trajectory after the first scattering.
It is then scattered again, and may return to the first black hole, and so on.
Since \( D\gg 2M \), we consider the scattering process of each black hole
separately and use formulas (\ref{eqDelta}) and (\ref{eqf}) to determine the
deflection angle due to each black hole as a function of the incident impact
parameter. 

By making the approximations mentioned above, we reduce the motion of light
in the two-black-hole space-time to a 2-d map, as has been done in
\cite{troll} to study general features of chaotic scattering. To do this, we make the further
assumption that the light rays have zero angular momentum in the direction of
the axial symmetry axis, on which lie the two black holes; the orbits are then
confined to a plane containing the two black holes. Due to the axial symmetry
of the system, the motions on all such planes are similar. Now suppose we have
a light ray escaping from one of the black holes with impact parameter \( b_{n} \)
and with an escaping angle \( \phi _{n} \) with respect to the symmetry axis,
as shown schematically in Figure 2. Since the black holes are considered to
be very far apart, the impact parameter \( b_{n+1} \) of the light ray with
respect to the other black hole is the segment \( l \) shown in Figure 2 (one
black hole can be considered to be ``at infinity'' as regards the other). We
use the convention that positive values of \( b \) means that the ray is directed
to the right side of the black hole, and rays with negative \( b \) are directed
to the left. From elementary geometry, we have \( l=b_{n}+D\sin \phi  \). The
deflection angle is given by \( \Delta (b_{n+1}) \). The map is then written
as:
\begin{equation}
\label{mapb}
b_{n+1}=b_{n}+D\sin \phi _{n};
\end{equation}
\begin{equation}
\label{mapphi}
\phi _{n+1}=\pi +\phi _{n}-\Delta (b_{n+1})\qquad \textrm{mod }2\pi .
\end{equation}
The angles \( \phi _{n} \) are measured counterclockwisely with respect to
each black hole; the first term in Eq. (\ref{mapphi}) comes from the change
in the angle's orientation necessary to take account of that. 

Consider the initial conditions \( b_{0}=b_{esc} \) and \( \phi _{0}=0 \).
Since \( \Delta (b_{esc})=\pi  \), we see from the above Equations that these
values of \( b \) and \( \phi  \) are a fixed point of the map. It corresponds
to the periodic orbit depicted in Figure 3a, which revolves around the black
holes, making a U-turn at each black hole and then heading towards the other.
Another periodic orbit is shown in Fig. 3b. This orbit is such that \( b_{n+1}=-b_{n} \)
and \( \phi _{n+1}=-\phi _{n} \). Inserting these conditions in Eqs. (\ref{mapb})
and (\ref{mapphi}), we find \( 2b_{0}=-D\sin \phi _{0} \) and \( \Delta (b_{0})=\pi +2|\phi _{0}| \)
(remember that the angles are defined modulus \( 2\pi  \)), with \( b_{0}>0 \).
\( \phi _{0} \) is given by the solution of the equation
\[
\Delta \left( \frac{D}{2}\sin |\phi _{0}|\right) =\pi +2|\phi _{0}|.\]

These are the simplest periodic orbits, but there are many others.

We observe that Equations (\ref{mapb}) and (\ref{mapphi}) are valid only as
long as \( b \) remains within the range \( b_{c}<b<b_{esc} \). If \( b \)
falls out of this interval, the ray either escapes or falls into one of the
black holes, and the iteration must be stopped.

\section{Analysis of the scattering map}

We now proceed to study in detail the map defined by Equations (\ref{mapb})
and (\ref{mapphi}). We begin by a direct numerical investigation of these equations.

In order to iterate Equations (\ref{mapb}) and (\ref{mapphi}) for given initial
values \( \phi _{0} \) and \( b_{0} \), we first have to be able to calculate
the deflection angle \( \Delta  \) for a given impact parameter \( b \). To
do this, we must begin by finding the ``perihelium distance'' \( P \) corresponding
to \( b \); this is done by solving the third-order equation (\ref{eqP}) for
\( P \). We use the well-known Newton-Raphson method, which guarantees a very
fast convergence\cite{nrec}. We then use Eqs. (\ref{eqQ}), (\ref{eqk}) and
(\ref{eqchi}) to calculate \( Q \), \( k \) and \( \chi  \), and we finally
substitute these quantities in Eqs. (\ref{eqDelta}) and (\ref{eqf}) to obtain
\( \Delta (b) \). The elliptical functions \( F \) and \( K \) are computed
by numerical routines found in \cite{nrec}.

Depending on its initial conditions, a light ray may either fall into one black
hole, fall into the other black hole, or escape towards infinity. The set of
initial conditions which leads to each of these outcomes is called the \emph{basin}
of that outcome. In our numerical iteration of the map (\ref{mapb},\ref{mapphi}),
we are interested in obtaining a basin portrait of the system. To do this, we
have to choose a set of initial conditions and iterate them to find out to which
basin they belong. Our choice is the one-dimensional set with \( \phi _{0}=0 \)
and an interval of \( b \). As we have seen in the previous Section, if \( |b|<b_{c} \),
the light ray always falls into the event horizon, and if \( |b|>b_{esc} \),
it always escapes. We thus choose the interval to be \( b_{c}<b<b_{esc} \).
We divide this segment into 5000 points, and iterate the map (\ref{mapb},\ref{mapphi})
for each of these initial conditions, recording the final outcome for each point:
if at any point in the iteration \( |b_{n}|<b_{c} \), this means that the light
ray falls into one of the black holes, and if \( |b_{n}|>b_{esc} \), it escapes
to infinity. We define the discrete-valued function \( g(b) \) to be 1 if the
orbit with initial conditions \( \phi _{0}=0 \), \( b_{0}=b \) falls into
one of the black holes, -1 if it falls into the other, and 0 if it escapes to
the asymptotically plane region. \( g(b) \) gives a picture of the intersections
of the three basins with the segment \( \phi _{0}=0 \), \( b_{c}<b<b_{esc} \). 

The result is shown in Fig. 4a, for \( D=15M \). We see that there are large
intervals in which \( g \) is constant, intercalated by ranges of \( b \)
where \( g \) varies wildly. If \( b_{0} \) lies within one of these latter
ranges, the final outcome of the light ray is highly uncertain. In Fig. 4b we
show a magnification of one of these regions. Except for the scale, it is very
similar to Fig. 4a. A further magnification is shown in Fig. 4c, again revealing
structure in small scales. We have obtained even further magnifications, which
are not shown here, and all show similar structures, down to the smallest scales
allowed by the numerical limitations. This shows that \( g \) has a fractal
dependency on \( b \). Notice that there are large intervals of \( b \) where
\( g \) is perfectly regular. These regular regions are mixed in all scales
with the fractal regions, where the outcome of a light ray is highly uncertain.
This sensitivity of the dynamics to the initial conditions is made precise with
the definition of the \emph{box-counting dimension}, which we now present briefly\cite{ott}.

We define the \emph{basin boundary} of the system to be the set of points (initial
conditions) such that all neighborhoods of these points have points belonging
to at least two different basins, no matter how small that neighborhood is.
The fractal nature of the basins shown in Fig. 4 results from a fractal basin
boundary\cite{ott}. It is not difficult to see that a fractal basin boundary
implies a fundamental uncertainty in the final outcome of an orbit. We now define
the box-counting dimension of the basin boundary, which gives a measure of this
uncertainty. Let \( b_{0} \) be a randomly chosen impact parameter in the interval
\( [b_{c},b_{esc}] \); we consider \( \phi _{0}=0 \) throughout for simplification.
Let \( f(\epsilon ) \) be the probability that there is a point of the basin
boundary lying within a distance \( \epsilon  \) from \( b_{0} \). In the
limit \( \epsilon \rightarrow 0 \), \( f \) generally scales with \( \epsilon  \)
by a power law. We thus write:
\begin{equation}
\label{scalf}
f(\epsilon )\propto \epsilon ^{1-d}.
\end{equation}
\( d \) is the box-counting dimension of the intersection of the basin boundary
with the one-dimensional section of initial conditions given by \( b\in [b_{c},b_{esc}] \)
and \( \phi _{0}=0 \). Clearly, we must have \( 0\leq d\leq 1 \). If the basin
boundary is regular, then \( d=0 \); fractal boundaries have \( d>0 \). \( f \)
can be interpreted as a measure of the uncertainty as to which basin the point
\( b \) belongs, for a given error \( \epsilon  \) in the initial condition,
which is always present in a real situation. For a regular basin boundary, \( f \)
decreases linearly with \( \epsilon  \). If we have a fractal boundary, however,
the power in (\ref{scalf}) is less than 1, and \( f \) decreases much more
slowly with \( \epsilon  \), which makes the uncertainty in the outcome much
higher than in the case of a regular boundary. Thus, \( d \) is a good measure
of the sensitivity to the initial conditions that results from a fractal basin
boundary, and since it is a topological invariant\cite{ott}, it is a meaningful
characterization of chaos in general relativity.

We calculate the box-counting dimension \( d \) numerically by using the method
we now explain\cite{ott}. We pick a large number of initial conditions \( b \)
randomly, and for each one of them we compute the map (\ref{mapb}-\ref{mapphi}),
finding out its outcome and therefore to which basin it belongs. We then do
the same thing to the two neighboring initial conditions \( b+\epsilon  \)
and \( b-\epsilon  \), for a given (small) \( \epsilon  \), for each \( b \).
If the three points do not belong to the same basin, \( b \) is labeled an
``uncertain'' initial condition. For a large number of initial conditions, we
expect that the fraction of uncertain points for a given \( \epsilon  \) approximates
\( f(\epsilon ) \). Calculating in this way \( f \) for several values of
\( \epsilon  \), we use Eq. (\ref{scalf}) to obtain \( d \) from the inclination
of the log-log plot of \( f \) versus \( \epsilon  \). Applied to the two-black-hole
map with \( D=15M \), this method gives \( d=0.17\pm 0.02 \). The error comes
from the statistical uncertainty which results from the finite number of points
used in the computation of \( f \). In our calculation, the number of initial
conditions was such that the number of ``uncertain points'' is always higher
than 200.

How does the fractal basin boundary arise from the dynamics of the map (\ref{mapb}-\ref{mapphi})?
In order to answer this question, we first observe that every point in the basin
boundary gives rise to orbits that neither escape nor fall into one of the black
holes (otherwise they would be part of one of the basins, which violates the
definition of the basin boundary); that is, the basin boundary is made up of
``eternal orbits'' which move forever around the two black holes. We need thus
to investigate these orbits to understand the formation of the basin boundary.

Consider the one-dimensional set of initial conditions parameterized by the
impact parameter \( b \) with \( \phi _{0}=0 \). We have seen that if \( |b|<b_{c} \)
the orbit falls into the event horizon of a black hole, and if \( |b|>b_{esc} \)
the orbit escapes. Thus, the points of the basin boundary belong to the interval
\begin{equation}
\label{interv}
|b|\in [b_{c},b_{esc}],
\end{equation}
which is actually two disjoint intervals, corresponding to positive and negative
values of \( b \). However, not all points in this interval are part of the
basin boundary, of course; in order to survive the next iteration of the map
(\ref{mapb}-\ref{mapphi}) without escaping or falling, the corresponding orbits
must be deflected in such a way that they reach the other black hole with an
impact parameter within the interval (\ref{interv}). From Fig. 5 we see that
for this to happen the orbits must be deflected by an angle \( \theta  \) in
the neighborhood of \( (2n+1)\pi +\alpha  \), and either \( (2n+1)\pi  \)
or \( (2n+1)\pi +2\alpha  \), depending on the previous deflection suffered
by the orbit; the angle \( \alpha  \) depends on the distance separating the
black holes. \( n \) is the number of turns the orbit makes around one of the
black holes before moving on to the other one. For each \( n \), there are
two intervals of the deflection angle \( \theta  \) for which the orbit survives
the next iteration without escaping or falling; these two intervals correspond
to the positive and negative values of \( b \) satisfying Eq. (\ref{interv}).
In the first iteration, the initial interval (\ref{interv}) is divided into
infinitely many pairs of sub-intervals, each pair labeled by the number \( n \)
of times the orbit circles the black hole. From Eq. (\ref{expon}), intervals
corresponding to large \( n \)'s decrease exponentially with \( n \). In the
next iteration, each of these sub-intervals are themselves divided into an infinite
number of intervals, and so on in the next iterations. In the limit of infinite
iterations, the set of surviving orbits is a fractal set with zero measure.
This set is the basin boundary, and its fractality is responsible for the complex
dynamics shown in Fig. 4. The two fractal regions on the right of Fig. 4a consist
of orbits whose first scattering has \( n=0 \), that is, they are deflected
by the black hole by \( \pi  \) and \( \pi +\alpha  \). The leftmost fractal
region in Fig. 4a is actually an infinite number of very small regions, corresponding
to orbits with \( n\neq 0 \); the scale of Fig. 4a is too large for them to
be distinguished. This gives us an indication that the orbits with \( n>0 \)
are a very small fraction of the basin boundary; we shall return to this later
in this Section.

Each orbit which is part of the basin boundary can be labeled by an infinite
sequence of symbols \( a_{1}a_{2}a_{3}\cdots  \) (including bi-infinite sequences
\( \cdots a_{-2}a_{-1}a_{0}a_{1}a_{2}\cdots  \)), where each symbol \( a_{m}=n_{m}(k_{m}) \)
gives the neighborhood of the deflection angle \( (2n_{m}+1)\pi +k_{m}\alpha  \)
after the \( m \)-th scattering. As an example, the periodic orbit shown in
Fig. 3a is represented by the sequence \( 0(0)0(0)0(0)\cdots  \). In general,
periodic orbits correspond to repeating sequences. However, not all sequences
are allowed. It is clear from Fig. 5 that a symbol \( n(0) \) must be followed
either by one of type \( m(0) \) or \( m(1) \), but it cannot be followed
by a symbol like \( m(2) \), that is, of the form \( m(k) \) with \( k=2 \).
Analogously, a symbol with \( k=1 \) of \( k=2 \) cannot be followed by one
with \( k=0 \). Even with these restrictions, however, there is an uncountable
set of non-repeating sequences which label orbits that are part of the basin
boundary. The uncountability of this set is a reflection of the fractal nature
of the boundary.

The basin boundary is the stable manifold of the invariant set, which is made
by orbits labeled by bi-infinite symbols \( \cdots a_{-2}a_{-1}a_{0}a_{1}a_{2}\cdots  \).
These are orbits that do not escape for both forward and backward iterations
of the map (\ref{mapb}-\ref{mapphi}).

It is important to observe that the basin boundary is fractal because the scattering
function \( \Delta (b) \) of the isolated black hole (\ref{eqDelta}) assumes
values higher than \( \pi  \), which makes it possible for orbits to be scattered
to both sides of the black hole, giving rise to the fractal basin boundary.
The scattering of particles by two fixed Newtonian mass points is immediately
seen to be regular, because Rutherford's scattering function does not assume
values higher than \( \pi  \). This is of course in accordance with the fact
that the fixed two-mass problem in Newtonian gravitation is integrable, since
the Hamilton-Jacobi equation of this system is separated in elliptical coordinates
\cite{whittaker}.

We have seen that after being scattered by a black hole, the light ray must
have an impact parameter lying on the interval (\ref{interv}) to belong to
the basin boundary. Since \( b_{c}=3\sqrt{3}M\approx 5.19615M \) and \( b_{esc}\approx 5.35696M \),
the impact parameter must belong to one of two intervals of length \( \Delta b=b_{esc}-b_{c}\approx 0.16081M \)
(the two intervals correspond to positive and negative impact parameters). In
our approximation, we have \( \Delta b\ll D \). Using this fact, we can approximate
the value of \( b \) in Fig. 5 by \( b_{esc} \), with an error of \( \Delta b \)
at worst, which means a fractional error of about \( \Delta b/b_{esc}\approx 0.03 \).
The distance \( L \) in Fig. 5 traveled by orbits which were deflected by \( (2n+1)\pi +\alpha  \)
or \( (2n+1)\pi +2\alpha  \) in the previous scattering is thus given by \( (L/2)^{2}+b_{esc}^{2}=(D/2)^{2} \),
that is,
\begin{equation}
\label{eqL}
L=\sqrt{D^{2}-4b_{esc}^{2}}.
\end{equation}
For the map (\ref{mapb}-\ref{mapphi}) to be well-defined, we must have \( D>2b_{esc} \).
Of course, this condition is satisfied in our approximation \( D\gg 2M \).
The angle \( \alpha  \) is calculated from Fig. 5 in this approximation, using
elementary geometry:
\begin{equation}
\label{eqalpha}
\sin \alpha =\frac{2b_{esc}}{D}.
\end{equation}

Consider a set of orbits with impact parameters filling the interval \( [b_{c},b_{esc}] \)
of length \( \Delta b \), which may have already suffered several previous
scatterings. The subsets of these orbits that survive the next scattering without
escaping nor falling in one of the black holes are sub-intervals in the neighborhood
of \( b_{n}^{k} \), where \( b_{n}^{k} \) are the values of the impact parameter
such that the orbit is deflected by an angle of \( (2n+1)\pi +k\alpha  \) (\( k=0 \),
1 or 2). They are solutions of the algebraic equation
\begin{equation}
\label{eqbnk}
\Delta (b_{n}^{k})=(2n+1)\pi +k\alpha ,
\end{equation}
with \( \Delta  \) given by Eq. (\ref{eqDelta}). The fractal regions of Fig.
4a are located around values of \( b \) given by Eq. (\ref{eqbnk}) with \( k=0 \)
and \( k=1 \). Depending on the previous deflections suffered by a given set
of trajectories, the values \( k=1 \) and \( k=2 \) must be used instead in
Eq. (\ref{eqbnk}), according to the rules of the symbolic dynamics we exposed
above.

Because the distance \( D \) is much larger than \( \Delta b \), the allowed
range in the deflection angle \( \delta \theta _{n}^{k} \) around \( (2n+1)\pi +k\alpha  \)
of an orbit such that it arrives at the other black hole with \( b \) in the
interval (\ref{interv}) is approximately:
\begin{equation}
\label{eqdtheta}
\delta \theta _{n}^{0}=\frac{\Delta b}{D};\quad \delta \theta _{n}^{1,2}=\frac{\Delta b}{L},
\end{equation}
where \( L \) is given by Eq. (\ref{eqL}). The length \( \Delta b_{n}^{k} \)
of the interval of surviving orbits around \( b_{n}^{k} \) is in this approximation
much smaller than \( \Delta b \): \( \Delta b_{n}^{k}\ll \Delta b \). We can
therefore approximate \( \Delta b_{n}^{k} \) by the first-order expression
\( \Delta b_{n}^{k}\approx \delta \theta _{n}^{k}/|\Delta '(b_{n}^{k})| \),
with \( \Delta '=d\Delta /db \). We define \( \lambda _{n}^{k}=\Delta b_{n}^{k}/\Delta b \)
as the fraction of the interval \( [b_{c},b_{esc}] \) (or \( [-b_{esc},-b_{c}]) \)
occupied by the surviving orbits with \( b \) around \( b_{n}^{k} \). We have:
\begin{equation}
\label{eqlambda}
\lambda _{n}^{0}=\frac{1}{D|\Delta '(b_{n}^{0})|};\quad \lambda _{n}^{1,2}=\frac{1}{L|\Delta '(b_{n}^{1,2})|}.
\end{equation}

Given a value of \( D \), it is easy to solve Eq. (\ref{eqbnk}) numerically
for \( b_{n}^{k} \), with \( L \) and \( \alpha  \) given by Eqs. (\ref{eqL})
and (\ref{eqalpha}). For \( D=15M \), we find \( L\approx 10.5M \) and \( \alpha \approx 0.7956 \).
By a direct numerical calculation of \( \Delta  \) and its first derivatives,
we find:

\begin{equation}
\label{b00}
b_{0}^{0}=b_{esc};\quad \Delta '(b_{0}^{0})=-5.863/M;\quad \Delta ''(b_{0}^{0})=38.30/M^{2};
\end{equation}
\[
b_{0}^{1}=5.26629M;\quad \Delta '(b_{0}^{1})=-13.35/M;\quad \Delta ''(b_{0}^{1})=202.5/M^{2};\]
\[
b_{0}^{2}=5.22729M;\quad \Delta '(b_{0}^{2})=-31.66/M;\quad \Delta ''(b_{0}^{2})=1030/M^{2};\]
\[
b_{1}^{0}=5.19643M;\quad \Delta '(b_{1}^{0})\approx -3600/M;\quad \Delta ''(b_{1}^{0})\approx 1.3\times 10^{7}/M^{2}.\]
Here \( \Delta ''=d^{2}\Delta /db^{2} \). The second derivative of \( \Delta  \)
will be used below. Notice that \( b_{n}^{0} \) is independent of \( D \).
Using Eq. (\ref{eqlambda}), we obtain:

\[
\lambda _{0}^{0}=0.01137;\quad \lambda _{0}^{1}=0.007134;\quad \lambda _{0}^{2}=0.003008;\]
\[
\lambda _{1}^{0}\approx 1.85\times 10^{-5}.\]

We see that \( \lambda _{1}^{0} \) is two or three orders of magnitude smaller
that \( \lambda _{0}^{k} \), and the other \( \lambda _{n}^{k} \)'s with \( n\geq 1 \)
are even smaller. From Eq. (\ref{expon}), they decrease exponentially with
increasing \( n \). These values show that for most purposes we can disregard
the contributions to the dynamics from deflections with \( n>0 \): the measure
of the set of orbits that make multiple turns around a black hole is negligible.
This approximation will be used extensively in the next Section.

Equations (\ref{eqbnk}-\ref{eqlambda}) are not exact, because the derivative
\( \Delta '(b) \) varies in the intervals \( \Delta b_{n}^{k} \). To estimate
the error, we use the fact that the \( \Delta b_{n}^{k} \) are small. The error
\( \delta \lambda _{n}^{k} \) in \( \lambda _{n}^{k} \) is then to first order:
\[
\delta \lambda _{n}^{0}=\frac{\Delta b_{n}^{0}}{D}\frac{\Delta ''(b_{n}^{0})}{\left[ \Delta '(b_{n}^{0})\right] ^{2}};\quad \delta \lambda _{n}^{1,2}=\frac{\Delta b_{n}^{1,2}}{L}\frac{\Delta ''(b_{n}^{1,2})}{\left[ \Delta '(b_{n}^{1,2})\right] ^{2}}.\]
Using \( \Delta b_{n}^{k}=\Delta b\lambda _{n}^{k} \), we get the fractional
error \( \delta \lambda _{n}^{k}/\lambda _{n}^{k} \):
\begin{equation}
\label{errlambda}
\frac{\delta \lambda _{n}^{0}}{\lambda _{n}^{0}}=\frac{\Delta b}{D}\frac{\Delta ''(b_{n}^{0})}{\left[ \Delta '(b_{n}^{0})\right] ^{2}};\quad \frac{\delta \lambda _{n}^{1,2}}{\lambda _{n}^{1,2}}=\frac{\Delta b}{L}\frac{\Delta ''(b_{n}^{1,2})}{\left[ \Delta '(b_{n}^{1,2})\right] ^{2}}.
\end{equation}
 The terms \( \Delta ''(b_{n}^{k})/\left( \Delta '(b_{n}^{k})\right) ^{2} \)
are of the order of 1, and the terms \( \Delta b/D \) and \( \Delta b/L \)
are much smaller than 1. Thus, the fractional errors in the values \( \lambda _{n}^{k} \)
given by the approximate formula (\ref{eqlambda}) are very small, of about
\( 0.01 \) for \( D=15M \). In the limit of large \( D \), we have \( L\approx D \),
and \( \delta \lambda _{n}^{k}/\lambda _{n}^{k}\sim 1/D \): the fractional
error is inversely proportional to the separation \( D \), and the approximation
(\ref{eqlambda}) gets better and better as \( D \) increases.

\section{The limit \protect\( D\rightarrow \infty \protect \)}

We now take the limit \( D\gg 2b_{esc} \). From Eqs. (\ref{eqalpha}) and (\ref{eqL}),
we have that in this limit \( \alpha \rightarrow 0 \) and \( L\rightarrow D \).
Eqs. (\ref{eqbnk}) and (\ref{eqlambda}) then imply that \( b_{n}^{k}\rightarrow b_{n}^{0}\equiv b_{n} \)
and \( \Delta b_{n}^{k}\rightarrow \Delta b_{n}^{0}\equiv \Delta b_{n} \):
the two intervals of surviving orbits of a given \( n \) come closer and closer
as \( D \) increases, and their lengths become the same in this limit. Because
of the approximate equality in the lengths, the magnification of each interval
by \( (\lambda _{n})^{-1} \) gives approximately the same set of intervals.
In other words, the fractal basin boundary is \emph{self-similar} in this approximation. 

The box-counting dimension of this self-similar set is given by the solution
of the transcendental equation\cite{ott}:
\[
2\sum _{n=0}^{\infty }(\lambda _{n})^{d}=1.\]

As we have seen in the previous Section, \( \lambda _{1} \) is many orders
of magnitude smaller than \( \lambda _{0} \). Therefore, it is a good approximation
to neglect terms with \( n>0 \) in the above expression, and \( d \) can then
be explicitly written as:
\begin{equation}
\label{eqd0}
d=\frac{\ln 2}{\ln (\lambda _{0})^{-1}}.
\end{equation}
From Eq. (\ref{eqlambda}), we have \( (\lambda _{0})^{-1}=D|\Delta '(b_{esc})| \),
and we get:
\begin{equation}
\label{eqd}
d=\frac{\ln 2}{\ln D+\beta },
\end{equation}
with \( \beta =\ln |\Delta '(b_{esc})| \). For large \( D \), the box-counting
dimension decays logarithmically with the distance, which is a very slow decay
law. Even for the large distances found in astronomy, \( d \) is still non-negligible.
Since the box-counting dimension is linked to the scattering cross-section \cite{levin},
the slow decay of \( d \) with \( D \) shown in Eq. (\ref{eqd}) could have
observational consequences.

\section{The case of different masses}

Now we consider the case of two black holes with different masses \( M_{a} \)
and \( M_{b} \). We restrict ourselves to the limit \( D\gg 2b_{esc}^{a},2b_{esc}^{b} \),
where \( b_{esc}^{a} \) and \( b_{esc}^{b} \) are the impact parameters corresponding
to deflections of \( \pi  \) of the two black holes. Because the two masses
are different, the ranges \( \Delta b_{a} \) and \( \Delta b_{b} \) of impact
parameters for surviving orbits are different, and therefore the allowed range
of scattering angles depends on which black hole the orbit is heading to. This
means that the `shrinking factors' \( \lambda _{0}^{a} \) and \( \lambda ^{b}_{0} \)
(given by Eq. (\ref{eqlambda})) depend on the black hole. This spoils the property
of self-similarity, which is a feature of the equal-masses case in the limit
\( D\rightarrow \infty  \). However, since the orbits are scattered alternately
by the two black holes, the \emph{square} of the scattering map defined by the
system is self-similar, in the limit \( D\rightarrow \infty  \). 

After two iterations, of each interval \( [b_{c},b_{esc}] \) there remains
four subsets of surviving orbits, all with size of approximately \( \lambda _{0}^{a}\lambda _{0}^{b} \)
(we are not considering orbits with \( n>0 \)). Each of these subsets gives
rise to four others after two further iterations, and so on. The box-counting
dimension of the surviving set is given by\cite{ott} \( 4(\lambda _{0}^{a}\lambda _{0}^{b})^{d}=1 \),
that is:
\[
d=-\frac{\ln 4}{\ln \left( \lambda _{0}^{a}\lambda _{0}^{b}\right) }.\]
Substituting Eqs. (\ref{eqlambda}) and (\ref{b00}) with \( M \) replaced
by \( M_{a} \) and \( M_{b} \), we obtain \( \lambda _{0}^{a} \) and \( \lambda _{0}^{b} \),
and we find:
\begin{equation}
\label{eqdab}
d=\frac{\ln 2}{\ln D+\beta +\ln \sqrt{\eta }},
\end{equation}
 where \( \eta =M_{a}/M_{b} \), \( \beta =\ln \Delta '(b_{esc}^{a}) \). The
difference between Eqs. (\ref{eqdab}) and (\ref{eqd}) is the constant term
\( \ln \sqrt{\eta } \) in the denominator of \( d \). If \( M_{a}=M_{b} \),
then \( \eta =1 \) and Eq. (\ref{eqdab}) is equal to Eq. (\ref{eqd}), as
of course it should. In the limit \( D\rightarrow \infty  \), \( d \) also
decays logarithmically with distance in this case, as in the case of equal masses.

\section{Conclusions}

In this article we have studied the chaotic behavior of light rays orbiting
a system of two non-rotating fixed black holes. We have assumed that the black
holes are sufficiently far away from each other, so that we could consider the
motion of the light rays to be the result of the action of each black hole separately.
Since the equations of motion of a light ray in the space-time of an isolated
black hole can be solved analytically, using this approximation we reduce the
motion of the massless test particle to a 2-dimensional map. Numerical integration
of this map showed the existence of a fractal basin boundary, with an associated
fractional box-counting dimension. In the limit of a large separation distance
\( D \) between the two black holes, we have been able to obtain an analytical
expression to the asymptotic value of the box-counting dimension \( d \). We
found that \( d\sim (\ln D)^{-1} \) for large \( D \); this result also holds
for different black hole masses.

\section*{Figure Captions}

\begin{description}
\item [Figure~1]The deflection function \( \Delta (b) \) for an isolated Schwarzschild
black hole.
\item [Figure~2]Construction of the scattering map.
\item [Figure~3]Two examples of periodic orbits of the scattering map.
\item [Figure~4]Portrait of the basins as a function of the impact parameter \( b \),
for \( \phi _{0}=0 \). The `basin function' \( g(b) \) is defined to be 1
if the orbit with initial conditions \( \phi _{0}=0 \), \( b_{0}=b \) falls
into one of the black holes, -1 if it falls into the other black hole, and 0
if it escapes to the asymptotically plane region. Even though the function \( g(b) \)
assumes only discrete values, the points are connected by straight lines for
better visualisation. Two successive magnifications of a small region of the
initial interval are pictured, showing the fractal dependency of \( g \) on
\( b \).
\item [Figure~5](a) Two possible types of scatterings for an orbit which was previously
scattered by \( (2n+1)\pi  \); (b) two scatterings for an orbit which was previously
deflected by \( (2n+1)\pi +\alpha  \) or \( (2n+1)\pi +2\alpha  \).
\end{description}
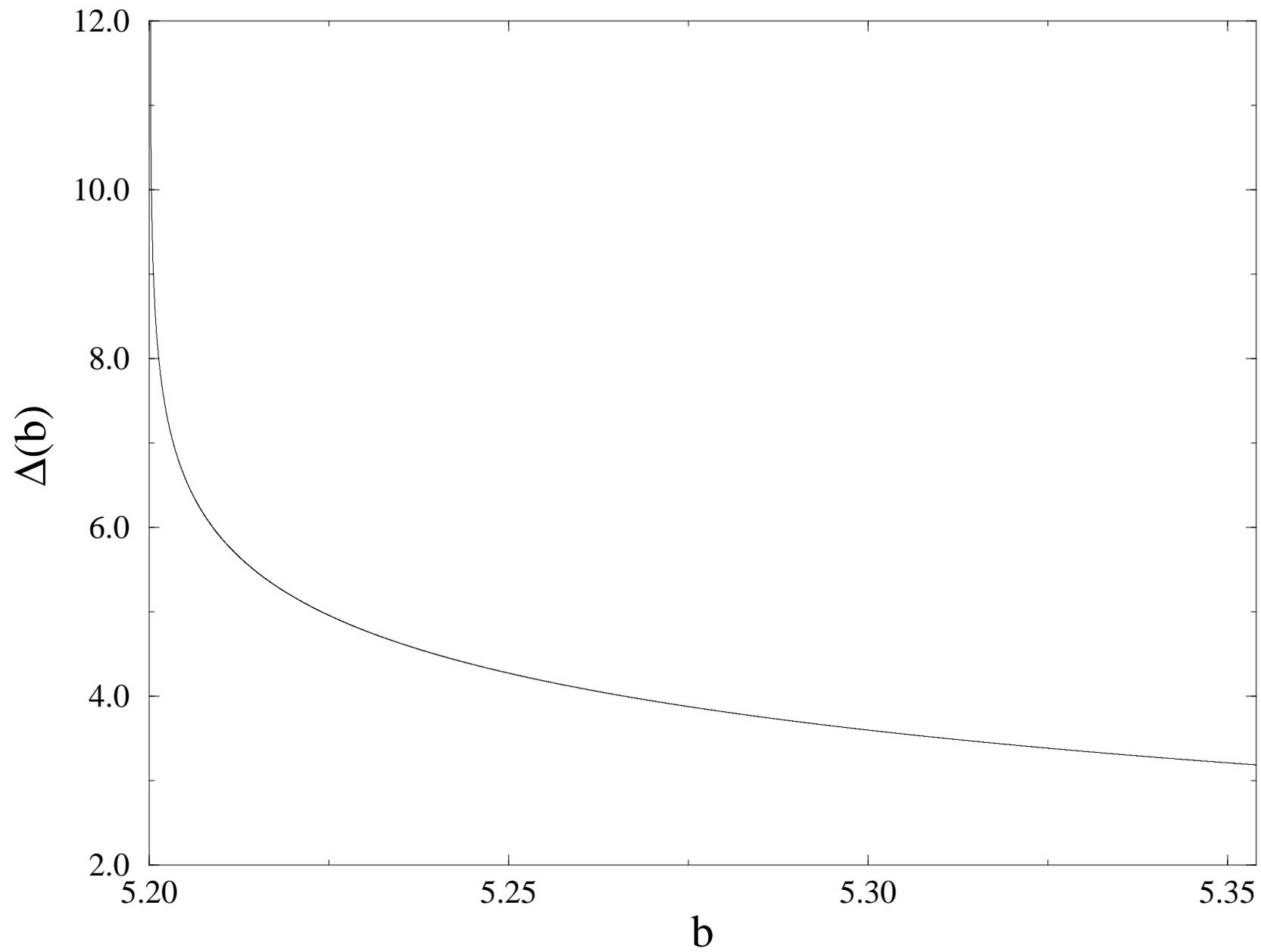
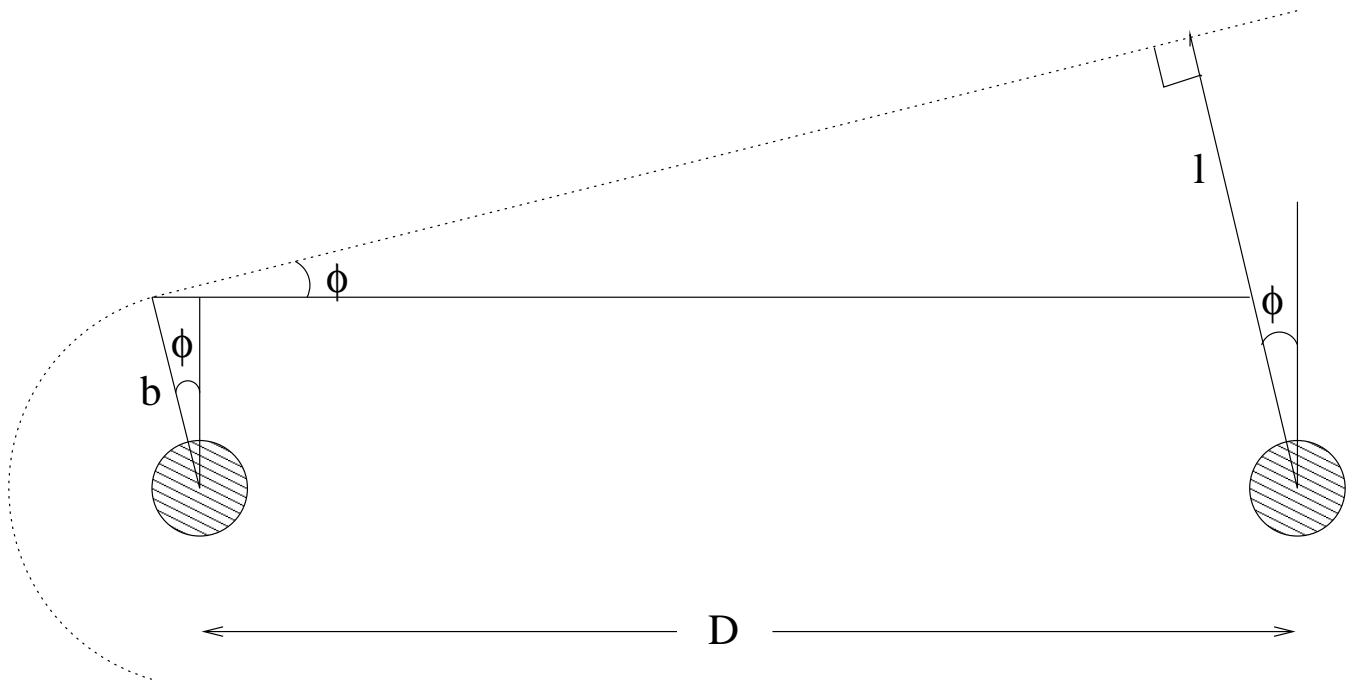
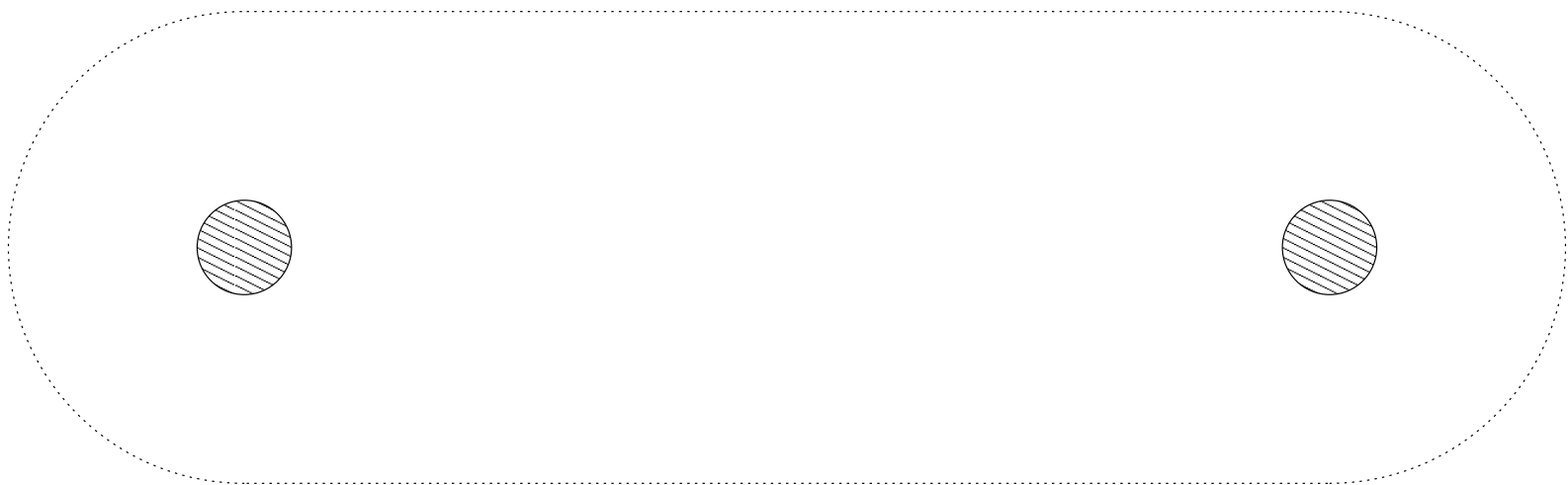
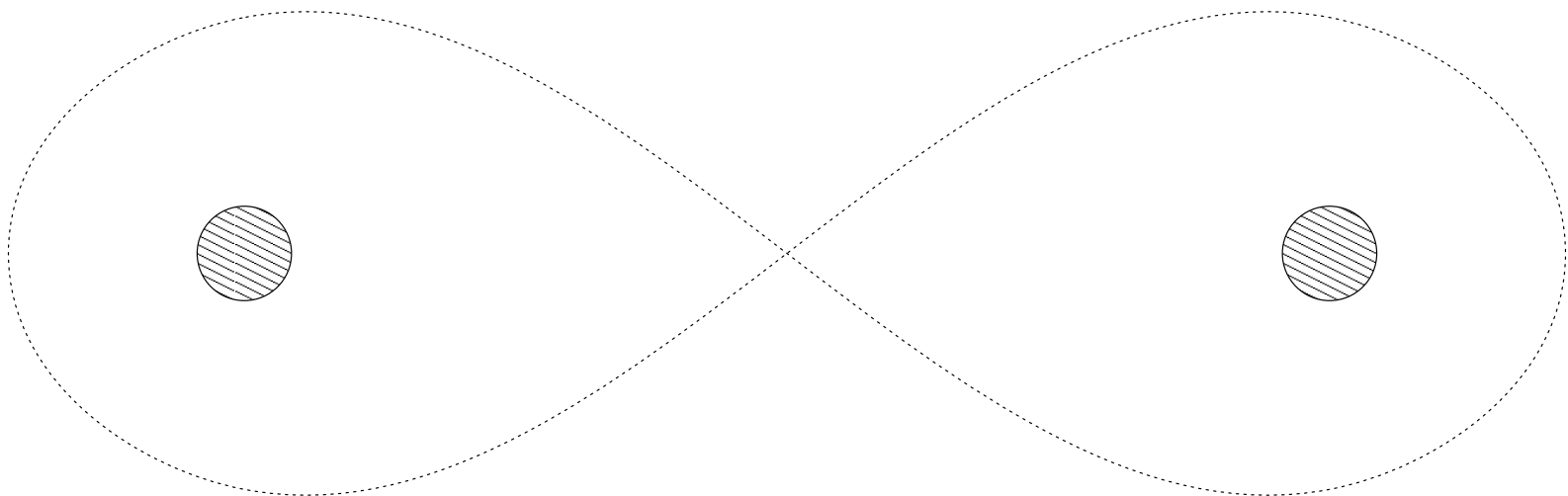
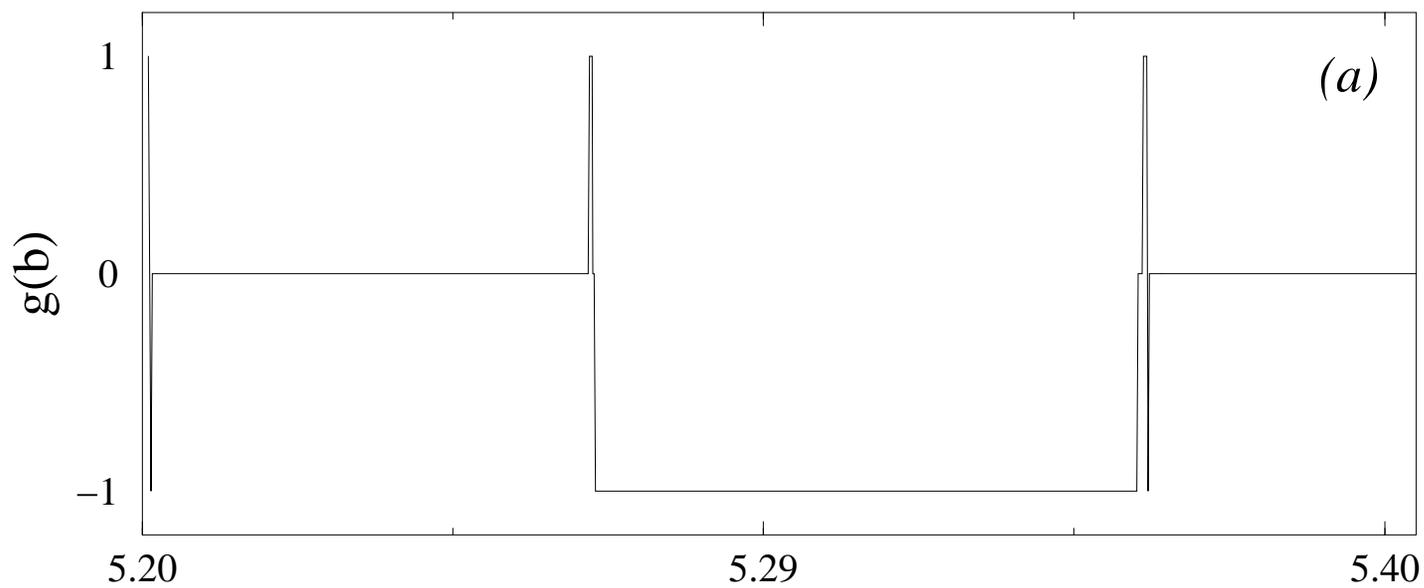
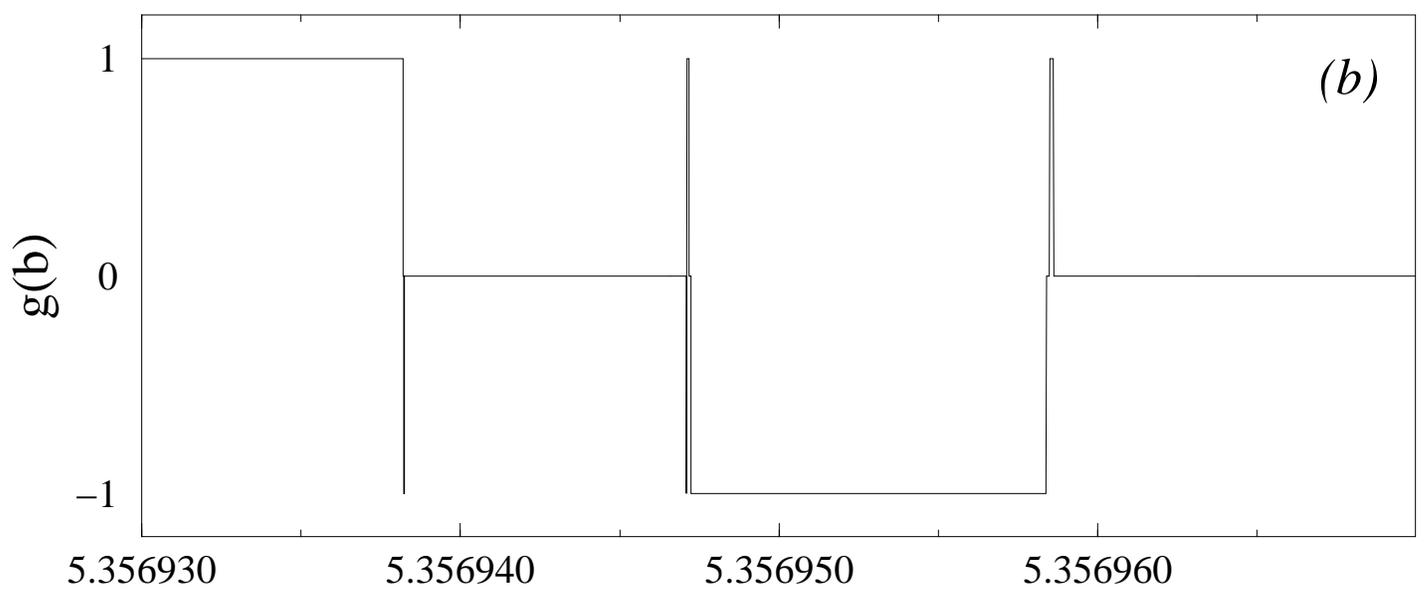
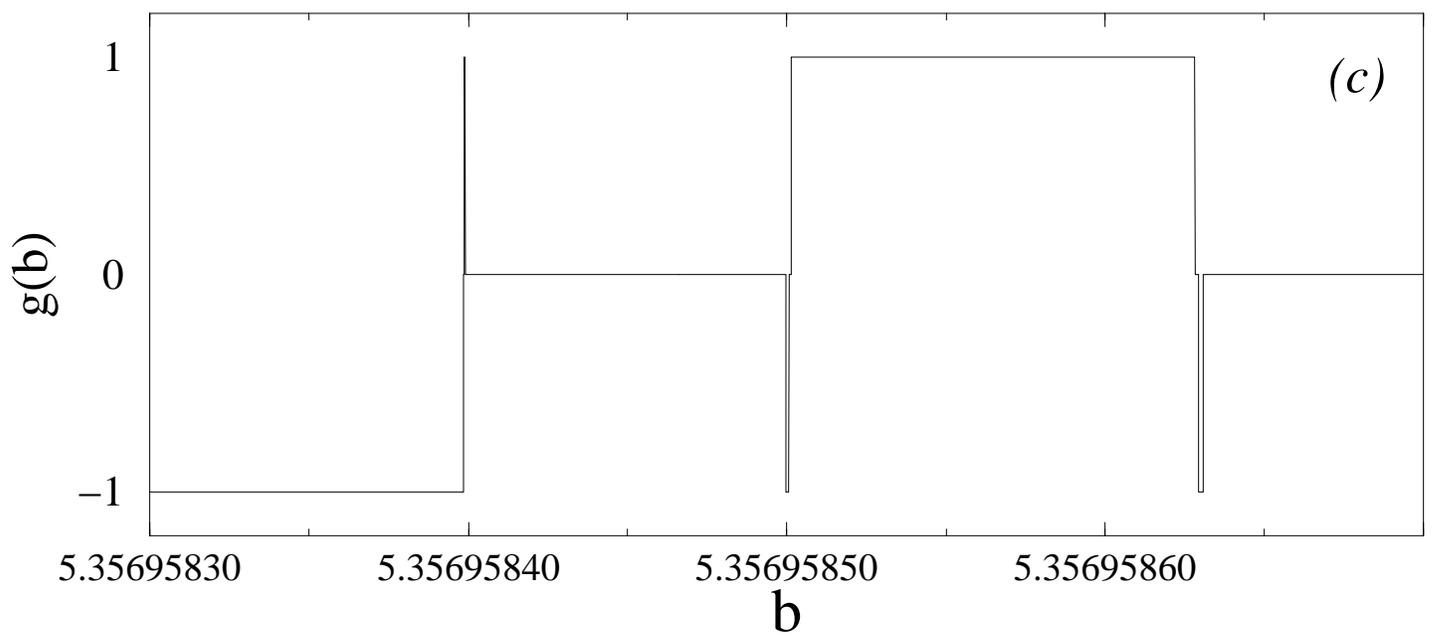
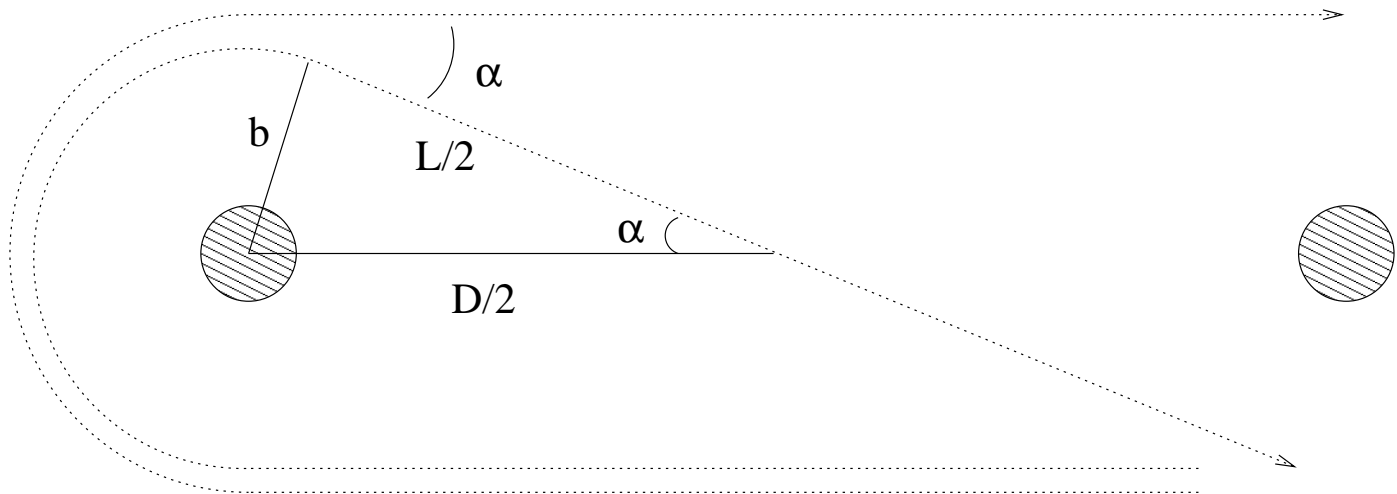
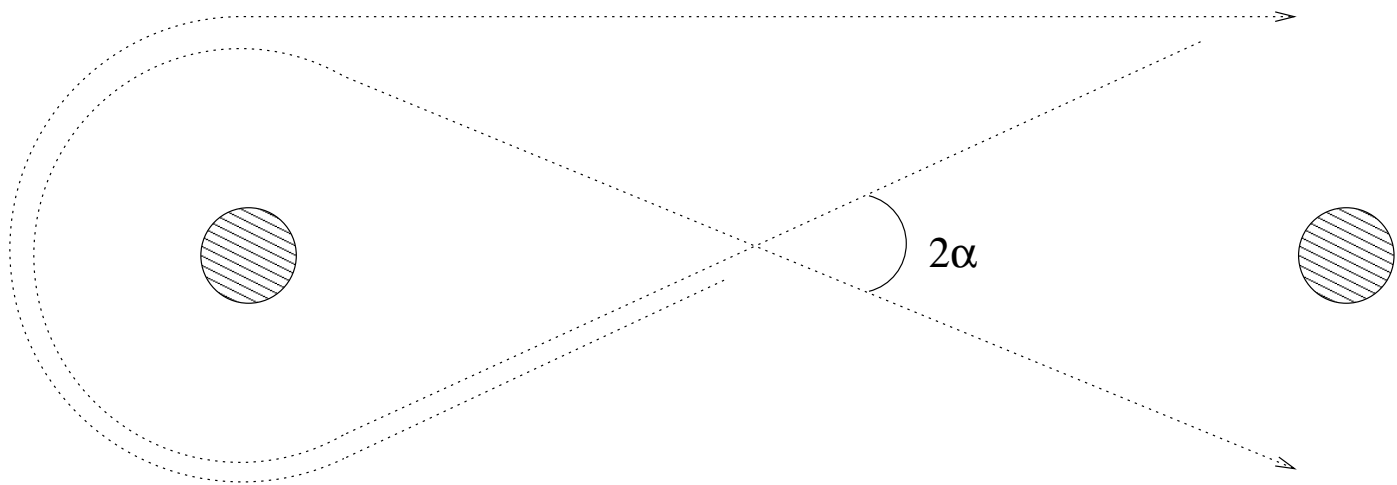
\end{document}